\documentclass[11pt,a4paper]{article}
\usepackage{amsfonts}
\usepackage{graphicx}
\usepackage{amsmath}
\usepackage[dvipdfm]{hyperref}
\usepackage{enumerate}
\RequirePackage{mathrsfs}
\RequirePackage[sc]{mathpazo}
\RequirePackage{wasysym}
\RequirePackage{setspace}

\makeatletter \@addtoreset{equation}{section}

\newcommand{\be}{\begin{equation}}
\newcommand{\ee}{\end{equation}}
\newcommand{\bea}{\begin{eqnarray}}
\newcommand{\eea}{\end{eqnarray}}
\begin{document}

\title{\bf Brane Realizations  of Quantum Hall Solitons and
Kac-Moody Lie Algebras }
\author{ A. Belhaj$^{1,2,5}$\thanks{belhaj@unizar.es}, A. ElRhalami$^{1,5}$,  N-E. Fahssi$^{1,3,5}$\thanks{fahssi@uh2m.ac.ma}, M. J. I. Khan$^{1,5}$, E. H. Saidi$^{1,5}$\thanks{h-saidi@fsr.ac.ma}, A. Segui$^{4}$\thanks{segui@unizar.es}\hspace*{-15pt} \\
\\
{\small $^{1}$Lab. Phys Hautes Energies, Mod\'elisation et
Simulation, Facult\'e
 des Sciences, Rabat, Morocco  } \\
{\small $^{2}$Centre National de l'\'Energie, des Sciences et des Techniques
Nucl\'eaires, Rabat, Morocco } \\
{\small $^{3}$D{\'e}partement de Math{\'e}matiques, Facult{\'e} des Sciences
et Techniques, Mohammedia, Morocco} \\
{\small $^{4}$Departamento de F\'isica Te\'orica, Universidad de Zaragoza,
E-50009-Zaragoza, Spain}\\
{\small $^{5}$Groupement National de Physique des Hautes Energies,
Si\`{e}ge focal, FSR, Rabat, Morocco }} \maketitle

\begin{abstract}
Using quiver gauge theories in  (1+2)-dimensions, we give brane
realizations of a class of Quantum Hall Solitons (QHS) embedded in
Type IIA superstring on the ALE spaces with exotic singularities.
These systems are obtained by considering two sets of wrapped
D4-branes on 2-spheres. The space-time on which the QHS live is
identified with the world-volume of D4-branes wrapped on a
collection of intersecting 2-spheres arranged as  extended Dynkin
diagrams of Kac-Moody Lie algebras. The magnetic source is given by
an extra orthogonal D4-brane wrapping a generic  2-cycle in the  ALE
spaces. It is shown as well that data on the representations of
Kac-Moody Lie algebras fix the filling factor of the QHS. In
case of finite Dynkin diagrams, we recover results on QHS with
integer and fractional filling factors known in the literature. In
 case of hyperbolic bilayer models, we obtain amongst others
filling factors describing holes in the graphene.
\end{abstract}

\thispagestyle{empty} \newpage \setcounter{page}{1} \newpage

\section{Introduction}

Three dimensional Quantum Hall Systems of condensed matter
physics can be engineered by low energy dynamics of D-branes
embedded in \textrm{Type II } superstrings \cite{BBST}--\cite{BS}.
In particular, a ten dimensional superstring picture of Quantum Hall
Solitons (QHS) in (1+2)-dimensions has been given in terms of a stack
of $K$ D6-branes and a spherical D2-brane  as well as dissolved
D0-branes in D2 and a stack of $N$ F-strings stretching between D2
and D6-branes \cite{BBST}. The world-volume of this D2-brane plays
the role of the 3-dimensional space-time, while the external source
of the magnetic charges is identified with the stack of the $K$
D6-branes which are placed perpendicular to the spherical D2-brane.
When the D6-branes cross the D2-brane, the Hanany-Witten effect
produces fundamental strings (F-strings) which are stretched between D2 and
D6-branes \cite{HW}. The F-strings ending on the D2-brane have an
interpretation in terms of the fractional quantum Hall particles
(Hall electrons), and they are charged under the U$(1)$ world-volume
gauge field associated with the D0-branes which behave as magnetic
flux quanta dissolved in the D2 world-volume.\\
In this paper, we elaborate  the brane  construction  of a class of QHS
that are embedded in Type IIA superstring  moving on asymptotically locally Euclidean
(ALE) spaces with non-trivial singularities \cite{FLRT,BS}. The (1+2)-dimensional QHS are engineered from a stack of
$n$ D4-branes wrapping intersecting 2-spheres arranged as Dynkin
diagrams of the Kac-Moody Lie algebras. These branes are coupled to
an external gauge field described by an extra D4-brane wrapping a
particular 2-cycle in ALE spaces playing the role of a magnetic
source. Its unwrapped directions are placed perpendicularly to the
uncompactified part of the word-volume of D4-branes on which a
U$(1)^{n}$ gauge theory lives. Among our results, we show that the
computation of the filling factors can be converted to a task of
studying representation theory of the Kac-Moody Lie algebras. In this work,
we examine models concerning quivers
associated with finite and hyperbolic sectors. Representations of
finite algebras classify models with integer, half-integer and
fractional filling factors. For the fundamental representation of finite
$A_{n}$ algebra for instance, we can recover celebrated filling
factor series. As for the case of bilayer hyperbolic models, we find
several subsequences of known series describing holes in the
graphene \cite{DSB}. Throughout this work, we give also multilayer
QHS models, based on the hyperbolic $H\widehat{A}_{n}$, with
positive, negative or zero filling factor values, depending on its
rank and vector charges.\\
This paper is organized  as follows. In section 2, we review briefly
some basic facts about Kac-Moody Lie algebras. In section 3, we
present  our  brane  construction  of QHS in (1+2)-dimensions from
Type IIA superstring on the ALE spaces  with deformed simply laced
singularities. In section 4, we compute the filling factors in terms
of weights and in section 5, we give explicit  models that are
associated with finite and hyperbolic sectors of Kac-Moody Lie algebras.
For the finite case, we first list $ADE$ quiver models with integer,
half-integer and fractional filling factors. Then, we examine  the
hyperbolic bilayered and multilayered systems with positive,
negative or zero filling factor values. Concluding discussion and
open questions are given in section 6.

\section{Generalized Cartan matrices}

Kac-Moody Lie algebras and their representations have been
extensively studied as they play a crucial role in many areas of
classical and quantum physics. They are behind the derivation of
several exact results in quantum field theory such as in $4D$
$\mathcal{N}=2$ conformal quiver gauge theories and also in the
compactification of superstring theory on Calabi-Yau manifolds by
using the geometric engineering method \cite{KMV}. They will be used
in this paper to develop new classes of quantum Hall solitons
embedded in six dimensional Type IIA superstring.\\
A nice way to introduce the Kac-Moody Lie algebras is in terms of
the generalized Cartan matrix \cite{V,Hu,ABS,ABS1} and the Dynkin
diagrams. For later use, let us give a brief review of some basic
facts about this method. The entries of the generalized Cartan
matrix satisfy
\begin{equation}
\begin{tabular}{llll}
$K_{ii}=2, $ &  & $K_{ij}<0,$ & if \; $K_{ij}=0\; $\ then $K_{ji}=0$.%
\end{tabular}%
\end{equation}%
According to Vinberg theorem on the classification of matrices, we
learn that one should distinguish three of generalized Kac-Moody Lie
algebras:

\begin{enumerate}
    \item \emph{Finite type}: $K_{ij}$ is a finite type
matrix if there exists a positive definite $u$ ($u_{i}>0;$
$i=1,2,...$) such that
\begin{equation}
K_{ij}^{(+)}u_{j}>0  \label{finite}
\end{equation}%
    \item \emph{Affine type}: $K_{ij}$ is an affine type matrix if there exists a
unique, up to a multiplicative factor, positive integer definite vector $u$ (%
$u_{i}>0;$ $i=1,2,...$) such that
\begin{equation}
K_{ij}^{(0)}u_{j}=0
\end{equation}%
    \item \emph{Indefined type: }$K_{ij}$ is an indefinite type matrix
if there exists a positive definite $u$ ($u_{i}>0;$ $i=1,2,...$)
such that
\begin{equation}
K_{ij}^{(-)}u_{j}<0.
\end{equation}%
\end{enumerate}
As the third sector is still an open problem in mathematics, we
shall restrict ourselves here below to its hyperbolic subset
obtained from the affine sector by adding an extra node to the
affine Dynkin diagram. The
upper indices $\pm $ and $0$ carried by the generalized Cartan matrices $%
K_{ij}^{\left( q\right) }$ introduced above are used to distinguish
the three sector which takes the following equivalent form:
\begin{enumerate}[(a)]
  \item Cartan matrices of finite dimensional Lie algebras satisfy $%
\hbox{det}\;K^{(+)}>0,$
  \item Cartan matrices of affine Kac-Moody Lie algebras are singular
since $\hbox{det}\;K^{(0)}=0,$
  \item  Cartan matrices of the hyperbolic subset obey the property $%
\hbox{det}\;K^{(-)}<0$.
\end{enumerate} Like in the finite dimensional case, we associate to any $n\times n$ Cartan
matrix $K_{ij}^{\left( q\right) }$ a generalized Dynkin diagram
which consists of $n$ nodes; and for each pair of nodes $(i,j)$ we
have $m_{ij}$ lines between them with $m_{ij}$ precisely given by
$K_{ij}$. This graph plays an important role in the study of quiver
gauge theories embedded into Type II superstrings compactified on
the K3 surface whose singularities are precisely classified by
Kac-Moody Lie algebras \cite{ABS,ABS1}. The deformation of these
singularities consists on blowing up the singular
point by a collection of intersecting 2-spheres with the intersection matrix $%
I_{ij}=-K_{ij}$.

\section{ Brane construction of  QHS}

Numerous  connections between string theory and QHS in
(1+2)-dimensions which were found during the last years. The first
of them was a construction of Bernevig, Brodie, Susskind and Toumbas
describing the QHE on a 2-sphere \cite{BBST} using a spherical
D2-brane and dissolved $N $ D0-branes moving on it. This
construction has been described in the introduction.\\
The ten dimensional string picture  of QHE in (1+2)-dimensions has
been extended to the compactification of Type IIA superstring theory
on the ALE spaces \cite{FLRT,BS}. In particular, one can
geometrically engineer at least two Type IIA brane realizations of
QHS \cite{BS}. The first realization uses D2 and wrapped D6-branes
wrapping the K3 surface, while the second one deals with only
D4-branes which we are interested in here. This approach is based on
the study of quiver gauge theories living on the world-volume of the
wrapped D4-branes on intersecting 2-spheres $S_{i}^{2}$ arranged as extended
Dynkin Diagrams. On the world-volume of theses wrapped branes lives
a U$(1)^{n}$ gauge symmetry in 3-dimensional space-time on which our
QHS will appear.\\
To couple the system to an external gauge field, one needs an extra
D4-brane wrapping a generic 2-cycle described by a linear
combination of $S_{i}^{2} $ as follows
\begin{equation}
\left[ C_{2}\right]=\sum_{i}q_{i}\left[S_{i}^{2}\right],
\end{equation}%
where $S_{i}^{2}$ denote a basis of $H_{2}(ALE,\mathbb{Z})$. A priori there are two
different ways in which the D4-brane is wrapped on each $S^{2}$. This can be
supported by the fourth homotopy group of $S^{2}$
\begin{equation}
\Pi _{4}(S^{2})=\mathbb{Z}_{2}.
\end{equation}%
Wrapping a D4-brane over such a geometry gives two possible D2-brane
configurations. Each one corresponds to an equivalent class of the
$\mathbb{Z}_{2}$ group. The wrapped D4-brane over a $S^{2}$ is
sensitive to this $\mathbb{Z}_{2}$ symmetry and, thus, it carries
two charges $\pm 1$. A general value of the charge $\pm q_i$ can be
obtained by wrapping a D4-brane $q_{i}$ times over a $S_{i}^{2}$ in
two possible orientations producing membranes which play the role of
the magnetic source in six dimensions. These membranes should be
placed perpendicularly to the uncompactified part of the word-volume
of D4-branes on which the $\mbox{U}(1)^{n}$ gauge theory lives. In
this realization of QHS, the $q_{i}$ charges play the same role as
the D6-brane ones in ten dimensions and can be interpreted as the
vector charges of
D-particles living on the space-time.  As in ten dimensional case \cite%
{BBST}, the missing spatial dimension is filled by
F-strings stretching between the orthogonal D4-branes (the ones on
which the $\mbox{U}(1)^{n}$ gauge theory lives and the one playing
the role of the magnetic source). The full brane system can be
described by the $\mbox{U}(1)^{n}$ Chern-Simons gauge theory with
the following action
\begin{equation}
S\sim \frac{1}{4\pi }\int \sum_{i,j}K_{ij}A^{i}\wedge
dA^{j}+2\sum_{i}q_{i}{\tilde{A}}\wedge dA^{i}.  \label{hd}
\end{equation}%
In this action, ${\tilde{A}}$ is an external $\mbox{U}(1)$ gauge
field and $A^{i}$ are dynamical gauge ones. $K_{ij}$ is a
\mbox{$n\times n$} matrix which can be related to the intersection
matrix of 2-spheres of ALE spaces \cite{FLRT,BS}. Following Wen-Zee
model \cite{WZ,wen}, $K_{ij}$ and $q_{i}$ are interpreted as order
parameters classifying the various QHS states and are related to the
filling factor via the relation
\begin{equation}
\nu =q_{i}K_{ij}^{-1}q_{j},  \label{factor}
\end{equation}%
which is invariant under the above $\mathbb{Z}_{2}$ symmetry. This
expression involves the inverse of the extended  Cartan matrix being
symmetric as required by simply laced singularities of the ALE
spaces.  We will show  that the computation of the filling factors
can be converted to a task of studying representation theory of
Kac-Moody Lie algebras. This means we will solve (\ref{factor}) in
terms of Dynkin geometry data on which QHS reside. In particular, we
will demonstrate that (\ref{factor}) can be solved using weight
equations. Then we illustrate our results for special choices for
the vector charge corresponding to known representations.

\section{ Algebraic solution of (\protect\ref{factor})}

In this section, we give an algebraic solution of (\ref{factor}) in
terms of weight space data of the corresponding algebra. This is not
a surprising result since the expression of the filling factors
involves the Cartan matrix of Kac-Moody Lie algebras. Thus, it is
natural to expect that representation theory will play an important
role for solving the equation (\ref{factor}). For simplicity, let us
restrict our self to finite dimensional algebras defined by
(\ref{finite}). They are classified in two types: simply laced $ADE$
algebras having a symmetric Cartan matrix and non simply laced
$BCFG$ ones having a non symmetric Cartan matrix. The last feature
is the main reason behind the complexity of the analysis of the blow
ups of the $BCFG$ singularities of the ALE surface. For this reason,
let us restrict our self to the simply laced case with
\begin{equation}
K_{ij}=(\alpha _{i},\alpha _{j})
\end{equation}%
where $\alpha _{i}$ are simple roots forming an orthonormal basis of
the root space. In this space, a weight vector $\Lambda $ can be
expressed as follows
\begin{equation}
\Lambda =\sum_{i}{\bar{\lambda}_{i}}\alpha _{i}
\end{equation}%
where $\bar{\lambda}_{i}$ are the coordinates of $\Lambda $ in the dual
basis. It turns out that, one can define, for any $\Lambda $, the so-called
Dynkin components $a_{i}$
\begin{equation}
\ a_{i}=\sum_{j}{\bar{\lambda}_{j}}\ K_{ij}
\end{equation}%
which are integer numbers \cite{group}. In Dynkin formalism, the scalar product
is given  by
the expression
\begin{equation}
(\Lambda ,\Lambda ^{\prime })=(\Lambda ^{\prime },\Lambda
)=\sum_{i,j}a_{i}^{\prime }K_{ij}^{-1}a_{j}.  \label{prod}
\end{equation}%
It follows that the vector norm reads as
\begin{equation}
(\Lambda ,\Lambda )=\sum_{i,j}{a_{i}}K_{ij}^{-1}a_{j}.
\end{equation}%
Now we take the vector charge $q_{i}$, which is the electromagnetic
charge carried by the extra D4-brane, to be the Dynkin labels
$a_{i}$. This  observation can be supported by a special choice of a divisor in
$H_{2}(ALE,\mathbb{Z})$ vector space. In this case, we should
consider a D4-brane in class \mbox{$[C_{2}]\in
H_{2}(ALE,\mathbb{Z})$} carrying magnetic charges under
$\mbox{U}(1)^{n}$ gauge symmetry described by
\begin{equation}
\lbrack C_{2}]=\sum_{i}a_{i}[S_{i}^{2}].
\end{equation}%
From representation theory, it follows that the expression of the filling
fractions (\ref{factor}) becomes
\begin{equation}
\nu =(\Lambda ,\Lambda ).  \label{norm}
\end{equation}%
Generally, one may have three situations classified by the signature of the
norm as given below
\begin{equation}\label{sign}
\begin{tabular}{lll}
(i) $\nu >0$, & (ii) $\nu =0$, & (iii) $\nu <0.$%
\end{tabular}%
\end{equation}
It should be interesting  to note that  the above results can have different
situations on weight lattices.  The most interesting are self-dual ones used in the construction
of the heterotic string theory. They can exist only in spaces with dimension $%
n=8d$ where $d$ is a positive integer. For $d=1$, there is just one lattice
called $E_{8}$, while in $d=2$ there are two special lattices which are $%
E_{8}\times E_{8}$ and $D_{16}$. In the case of $d=3$ there is a
special lattice called Leach lattice which has a minimal norm equal
4. Beside self-dual cases, there exit also integer-valued lattices
with minimal norm equal 1 and 2 which are called respectively odd
and even lattices.\\
In the sequel, we will adopt the solution (\ref{norm}) in our illustrating
examples.

\section{Explicit quiver  models}

In this section, we apply to stringy QHS some ideas familiar from
the study of the extended Cartan matrix associated with  quiver
gauge theories  to compute the filling factors. In particular, we give
some illustrating models for simply laced  finite and hyperbolic
Kac-Moody Lie algebras. For these algebras, the quadratic form
(\ref{factor}) can be written as follows
\begin{equation}
\nu=\sum_{i}K_{ii}^{-1}a_i^2+2\sum_{i<j} K_{ij}^{-1}a_ia_j.
\end{equation}

\subsection{ Finite quiver models}

Here, we consider QHS associated with finite dimensional simply
laced algebras of type $A_n$, $D_n$ and $E_s$ $(s=6,7,8$) for which
all roots have the same length squared. In particular, we describe
the $A_n$ case in detail. The results for the other algebras may be
obtained without difficulty. To that purpose, we  first collect  some
useful information. We stress that generally the determinant of the
Cartan matrix of all these series is always positive \cite{Hu}. They
are given by
\begin{equation*}
A_n : n+1\; ;\qquad D_n : 4\; ; \qquad E_6 : 3\; ; \qquad E_7 : 2 \;
; \qquad E_8: 1.
\end{equation*}
In the residue  of this section, we will give some concrete examples to
illustrate our results for integer and fractional values for the $\nu$%
-factor.

\subsubsection{$A_n$ quivers}

Consider a $\mbox{U}(1)^n$ quiver gauge theory with finite  $A_n$
Cartan matrix. In the context of string theory, this model appears
as the world-volume of $n$
D4-branes wrapping separately $n$ \mbox{2-cycles} of the deformed $\mathbb{C}%
^2/\mathbb{Z}_{n+1}$ singularity. The K\"{a}hler deformation of this geometry is
obtained by blowing up the singular point using $n$ intersecting 2-spheres, %
$\left(S_i^{2}\right)$, arranged as follows
\begin{equation}
\mbox{
         \begin{picture}(20,30)(70,0)
        \unitlength=2cm
        \thicklines
    \put(0,0.2){\circle{.2}}
     \put(.1,0.2){\line(1,0){.5}}
     \put(.7,0.2){\circle{.2}}
     \put(.8,0.2){\line(1,0){.5}}
     \put(1.4,0.2){\circle{.2}}
     \put(1.6,0.2){$.\ .\ .\ .\ .\ .$}
     \put(2.5,0.2){\circle{.2}}
     \put(2.6,0.2){\line(1,0){.5}}
     \put(3.2,0.2){\circle{.2}}
     \put(-1.2,.15){$A_{n}:$}
  \end{picture}
}  \label{ordAk}
\end{equation}
The elements of the corresponding Cartan matrix are given by $K_{ij}=2\delta_{ij}-%
\delta_{i\, j+1}-\delta_{i+1 \,j}$, from which the inverse matrix can be
evaluated to be
\begin{equation*}
K^{-1}_{ij}=\frac{i(n-j+1)}{n+1}, \qquad (i \leq j).
\end{equation*}
Thus, the filling factor (\ref{factor}) reduces to the following
quadratic form
\begin{equation}
\label{an}
\nu(A_n)=\frac{1}{n+1}\left(\sum_{i=1}^n i(n-i+1) a_i^2+2\sum_{i<j}
i(n-j+1)a_ia_j\right).
\end{equation}
For the sake of illustration we shall now calculate this
$\nu$-factor for some irreps of $A_n$ algebras. The simplest ones
are the so-called basic representations whose Dynkin components are
zeros except one entry which is equal 1. In the quantum Hall
literature, this example is related to single layer states. If we
take $a_i=\delta_{i p}$ for some $p =1, \ldots, n$, then equation
(\ref{an}) reduces to
\begin{equation}
\nu(A_n)=\nu_{p,n}=\frac{p(n+1-p)}{n+1}
\end{equation}
exhibiting the obvious symmetry $%
\nu_{p,n}=\nu_{n+1-p,n}$. In the fundamental representation (where $p=1$ or $n$), we find $%
\nu_{1,n}=1-\frac{1}{n+1}$  which coincides with known filling fractions in the
literature \cite{wen}. In table \ref{tab1}, we list the values of $\nu$
for the fundamental, the adjoint and the $\rho$-representation (that is
irreps whose highest weight is the Weyl vector $\rho$ which is half sum of
positive roots).

\begin{table}[htbp]
\centering
\begin{equation*}
\begin{tabular}{llll}
\hline
Irreps & Vector charge & $\nu$ & Comments \\ \hline\hline
fundamental & $(10 \ldots 00)$ or $(00 \ldots 01)$ & $\frac{n}{n+1}$ &
fractional \\
adjoint & $(10 \ldots 01)$ & 2 & for all $n$ \\
$\rho$-representation & $(11 \ldots 11)$ & $\frac{1}{2}{\binom{n+2 }{3}}$ & $%
\begin{array}{ll}
\hbox{half-integer} & \hbox{if}\; \; n \equiv 1 \mod 4 \\
\hbox{integer} & \hbox{otherwise} \\
&
\end{array}%
$ \\ \hline
\end{tabular}%
\end{equation*}%
\caption{The $\protect\nu$-factor associated with some irreducible
representations of $A_n$.}
\label{tab1}
\end{table}

At this stage, two points are worthy of notes

\begin{itemize}
\item If we fix the normalization of the inner product (\ref{prod}) to be
such that the highest weight of the adjoint representation has norm 1, then,
in the fundamental representation of $A_{2n}$ we have $(\Lambda,\Lambda)=%
\frac{n}{2n+1}$, which coincides with a subsequence of the celebrated Jain's
series. This series can besides be recovered as an exact value of the $\nu$%
-factor (\ref{factor}) if we have in the action (\ref{hd}), $2 K_{ij}$
instead of $K_{ij}$  \cite{BFSS} (or more generally $K_{ij}+K_{j\,i}$ in place of $K_{ij}$
if we wish to cover non symmetric Cartan matrices).

\item The $\nu$-factor is independent of the rank $n$ in the adjoint
representation. This fact is general for vector charges of the form $%
a_i=\delta_{i p}+\delta_{i\, n+1-p}$. We find in this case $\nu=2p$.
\end{itemize}
\subsubsection{$DE$ quivers}
The results for $DE$ quiver gauge models can straightforwardly be
obtained
in similar way. We just display, in the end of this subsection, in Table \ref%
{tab2} the values of the $\nu$-factor for simple irreps, which are
representations from which all other irreps can be constructed by
tensor products. \begin{table}[htbp] \centering
\begin{tabular}{lll}
\hline
& Dynkin designation & $\nu$ \\ \hline\hline
$D_n$ & $(10 \ldots 00)$ & 1 \\
& $(00 \ldots 01)$ or $(0 \ldots 010)$ & $n/4$ \\
$E_6$ & $(100000)$ or $(000010)$ & 4/3 \\
$E_7$ & $(0000010)$ & 3/2 \\
$E_8$ & $(00000010)$ & 2 \\ \hline
\end{tabular}%
\caption{The $\protect\nu$-factor associated with simple irreducible
representations of $D$ and $E$ algebras.}
\label{tab2}
\end{table}

\subsection{Hyperbolic  quiver models}

In this section, we consider indefinite quiver gauge theories
describing multilayers states in QHS. In this case, we will see
that the filling factor (\ref{factor}) can take positive, negative
or zero values, depending on the vector charges and the rank of the
algebras. To start, recall first that there is no full
classification for indefinite Cartan matrices and very little is
known about the corresponding  representation theory. For this
reason, we will not refer to representation theory and  we will
restrict our self to quiver gauge theories with a hyperbolic Cartan
matrix (matrices with a single negative eigenvalue and all the other
positive). Recall that hyperbolic Kac-Moody Lie algebras are by
definition Lorentzian Kac-Moody Lie algebras with the property that
cutting any node from their Dynkin diagrams leaves one with a Dynkin
diagram of affine or finite type.

The general study is beyond the scope of the present work, though we
will consider two explicit examples. The first example corresponds
to the case of two layers QHS, while the second one will be
associated with multilayer systems.

As mentioned, the starting example concerns a bilayer system corresponding
to $\mbox{U}(1)\times \mbox{U}(1)$ quiver gauge theory with the following Cartan matrix
\begin{equation*}
K_{ij}(\mbox{U}(1)\times \mbox{U}(1))=\begin{pmatrix}
2 & -k \\
-k & 2%
\end{pmatrix}
\end{equation*}%
In order to get a hyperbolic Dynkin geometry, one needs to require
\begin{equation}
4-k^{2}<0.
\end{equation}%
From string theory point of view, this model can be obtained from two
D4-branes wrapping on two intersecting 2-spheres $S_{i}^{2}$ with the
following  intersection constraint
\begin{equation}
S_{i}^{2}.S_{j}^{2}=-K_{ij}(\mbox{U}(1)\times \mbox{U}(1)).
\end{equation}%
Evaluating (\ref{factor}) for the charges $%
a_{i}=(1,1)$ yields
\begin{equation}
\nu =\frac{2}{2-k}.
\end{equation}%
This is a general expression containing some known series used in the study
of trial functions of QHE in the graphene. These kind of functions have been
proposed first by Laughlin \cite{L} and they have been generalized by
Halperin in order to understand multi-components with SU$(K)$ internal
symmetries \cite{H}. Based on Halperin ideas on the study of trial wave
functions $(m_{1},m_{2},m_{3})$ of QHE, the results we obtain here may
correspond to holes in the graphene structure. In particular, putting $k=2m+1
$, we recover a $\mbox{U}(1)\times \mbox{U}(1)$ quiver gauge theory of bilayers QHS with
the following filling fraction
\begin{equation}
\nu =\frac{2}{1-2m}.
\end{equation}%
The opposite value ($\nu =\frac{2}{2m-1}$) has been obtained in the study of
$(m,m,m-1)$ trial wave functions given in \cite{PGR}. One may also get $\nu
=-\frac{1}{m}$ by taking $k=2m+2$ and it corresponds to $(m,m,m-1)$
Laughlin's wave functions   describing holes. We note that one can
reproduce integer values like $\nu =-1$ and $\nu =-2$ by taking $k=3$ and $%
k=4$ respectively. These values has been dealt with to  study non
equilibrium breakdown of QHE in graphene \cite{SD}.\\
The next example of hyperbolic Kac-Moody Lie algebras that we give is related  the
over-extensions of affine ones. A simple situation is to add one extra node
to the Dynkin graph of the $\widehat A_n$ affine Kac-Moody Lie algebra. In
geometric Dynkin language, this corresponds to a particular hyperbolic
Dynkin diagram. The derivation of such a hyperbolic geometry is based on the
same philosophy one uses in the building of the affine Dynkin diagrams from
the finite ones by adding a node. In other words, by cutting this node in
such a hyperbolic Dynkin diagram, the resulting sub-diagram coincides with
the Dynkin graph of the $\widehat A_n$ affine Kac-Moody Lie algebras. We
refer to this Dynkin graph as $H\widehat A_n$
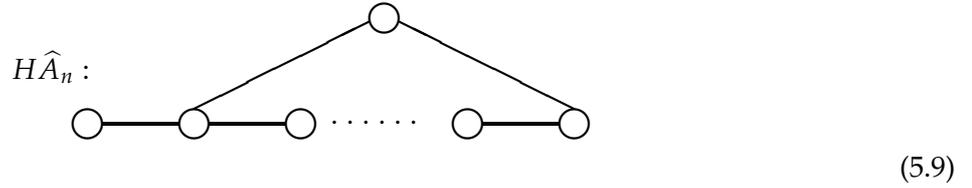
\begin{equation}
\mbox{
         \begin{picture}(20,90)(50,-20)
        \unitlength=2cm
        \thicklines
      \put(-1.2,.3){$H\widehat{A}_n:$}
\put(-0.7,0){\circle{.2}}
     \put(-0.6,0){\line(1,0){.5}}
    \put(0,0){\circle{.2}}
     \put(.1,0){\line(1,0){.5}}
     \put(.7,0){\circle{.2}}
     \put(.9,0){$.\ .\ .\ .\ .\ .$}
     \put(1.8,0){\circle{.2}}
     \put(1.9,0){\line(1,0){.5}}
     \put(2.5,0){\circle{.2}}
     \put(0,.1){\line(2,1){1.15}}
     \put(1.25,.7){\circle{.2}}
     \put(2.5,.1){\line(-2,1){1.15}}
  \end{picture}
}  \label{affAk}
\end{equation}%
The Cartan matrix of this algebra reads as
\begin{equation}
K_{i j} = 2\delta_{i,j}-\delta_{i,j+1}-\delta_{i+1,j} -\delta _{i, n+2}
\delta _{2, j} -\delta _{2, i} \delta _{j, n+2} \quad i,j = 1, \ldots n+2 ,
\end{equation}
from which we derive, rather laboriously, the inverse matrix elements (for $%
i \leq j$)
\begin{eqnarray}
K^{-1}_{1 \,j} &=& \left\{%
\begin{array}{ll}
0, & \hbox{if} \quad j=1 \\
-1, & \hbox{if} \quad j > 1 \\
\end{array}%
\right.  \notag \\
K^{-1}_{i \,j} &=& -\frac{1}{n+1}\left(i j+(4-i) n-3 i-2 j+8\right).
\end{eqnarray}
 This expression shows
that, interestingly, $\hbox{det}\,K(H\widehat A_n)=-(n+1)$. Adding
two nodes to the Dynkin diagram of the $A_n$ finite Kac-Moody Lie
algebra, as shown in the above figure, makes that the determinant of
the Cartan matrix is multiplied by -1. Mathematically, it would be
instructive to examine this property for more general
over-extensions. For the present case, (\ref{factor}) can be cast in
the following form
\begin{multline}
\nu(H\widehat A_n)= -\frac{1}{n+1}\sum _{i=2}^{n+2} \left(i^2+(4-i) n-5
i+8\right)a_i^2\, - \\
\frac{2}{n+1} \sum _{j=3}^{n+2} \sum _{i=2}^{j-1} (i j+(4-i) n-3 i-2 j+8)\,
a_i a_j -2 a_1 \sum _{j=2}^{n+2} a_j .
\end{multline}
Numerical investigations show that, for any vector
charge, the sign of $\hat{\nu}_{p,n}$ is not constant and vanishes
for special values of $n$ and $a_i$, covering the three signatures
of (\ref{sign}). This reveals the Lorentzian nature of the weight
space for hyperbolic algebras. In Table (\ref{tab3}), we will
collect some remarkable values of $\nu$.

For the case where $a_i=\delta_{i p}$ (for some $p =1, \ldots, n$),
the filling factors  read as
\begin{equation}
\nu(H\widehat A_n)=K^{-1}_{pp}=-\frac{1}{n+1} \left(
p^2-(n+5)p+4n+8\right):= \hat{\nu}_{p,n},
\end{equation}
for $p=2,\ldots,n+2$.
It is easy to show that the following symmetry relation is fulfilled
\begin{equation*}
\hat{\nu}_{p,n}=\hat{\nu}_{5+n-p,n}
\end{equation*}
which may  reflect a symmetry in the  corresponding lattice weight space. It can be also shown
that $\hat{\nu}_{p,n}$ is zero only for the values $(p,n)=(5,8),(8,8),(6,7)$. Here, we list the following result
\begin{table}[htbp]
\centering
\begin{equation*}
\begin{tabular}{lll}
\hline
Vector charge & $\nu$ & Comments \\ \hline\hline
$(10 \ldots 00)$ & 0 & for all $n$ \\
$(01 \ldots 00)$ & -2 & for all $n$ \\
$(000010 \ldots 00)$ or $(00 \ldots 0100)$ & $(n-8)/(n+1)$ & integer for $n
=2,8 $ \\
$(0000010 \ldots 0)$ or $(00 \ldots 1000)$ & $2 (n-7)/(n+1)$ & integer for $%
n =1, 3, 7, 15 $ \\
$(0010 \ldots 01)$ & -6 & for all $n$ \\
$(11 \ldots 11)$ & $\frac{1}{12}(n-24)(n+1)(n+2)$ & $%
\begin{array}{ll}
\hbox{half-integer} & \hbox{if}\; \; n \equiv 1 \mod 4 \\
\hbox{integer} & \hbox{otherwise} \\
\end{array}%
$ \\ \hline
\end{tabular}%
\end{equation*}%
\caption{The $\protect\nu$-factor associated with  $H\widehat A_n$.}
\label{tab3}
\end{table}

\bigskip \noindent From  the significant observation of the above table we conclude that the value -6 in Table \ref{tab3} is a
particular case of the following more general fact. If
$a_i=\delta_{i p}+ \delta_{i q}$ such that $p+q=n+5$, ($p \geq 3$),
then $\nu(H\widehat A_n)=2(p-6)$  clearly vanishes for the
vector charge $(0000010 \ldots 01000)$. It should be instructive to
understand these properties of hyperbolic QHS.

\section{Discussions and Open questions}

In this paper we have discussed brane realizations of QHS associated
with extended Cartan matrices of Kac-Moody Lie algebras. The models
considered here are described by abelian quiver gauge theories
obtained from the compactification  of Type IIA superstring theory
on the ALE spaces with deformed simply laced singularities. In
particular, we have shown that the computation of the filling
factors  can be converted to a task of studying representation
theory of Kac-Moody Lie algebras. For some concrete models involving
intersecting D4-branes on two-spheres arranged as hyperbolic Dynkin
diagrams with two nodes, we have obtained values for the filling
factors  which coincide with several subsequences of known series
used in graphene. Concerning the finite  quiver gauge models, we
have found several QHS with integer, half-integer and fractional
filling factors (with  odd-denominators). In particular, one may
recover the celebrated Jain's series from the fundamental
representation of $A_{2n}$ algebra.\\
Our work opens up for further studies. One interesting problem is to
complete the analysis by considering non abelian quiver gauge
theories. It would therefore be of great interest to try to extract
information on the exact connection between the graphene and
hyperbolic sector of the Kac-Moody Lie algebras. We believe that
this observation deserves to be studied further. We hope to report
elsewhere on these opens questions.

\emph{Acknowledgments.} This work is supported by the program
Protars III D12/25. AS is  supported by CICYT (grant FPA-2006-02315
and grant FPA-2009-09638) and DGIID-DGA (grant 2007-E24/2). We thank
also the support by grant A/9335/07 and A/024147/09. AB  would like
to thank  M. Asorey for discussions and scientific help and also
Departamento de F\'isica Te\'orica, Universidad de Zaragoza for kind
hospitality.

\end{document}